\documentclass[conference]{IEEEtran}
\IEEEoverridecommandlockouts
\usepackage{cite}
\usepackage{soul}
\usepackage{amsmath,amssymb,amsfonts}
\usepackage{amsthm}
\usepackage{graphicx}
\usepackage{textcomp}
\usepackage{xcolor}
\usepackage{multicol}
\usepackage{cuted} 
\usepackage{arydshln}
\usepackage{stfloats}
\usepackage{hyperref}

\usepackage{algorithm}
\usepackage[noend]{algpseudocode}
\allowdisplaybreaks
\newtheorem{theorem}{Theorem}

\newtheorem{remark}{Remark}

\newcommand{\btt}[1]{{\fontfamily{lmtt}\selectfont #1}\normalfont}

\definecolor{niceblue}{rgb}{0.125, 0.406, 0.852}

\hypersetup{
    colorlinks=true,
    linkcolor=niceblue,
    filecolor=magenta,      
    urlcolor=niceblue,
    citecolor=niceblue}

\def\BibTeX{{\rm B\kern-.05em{\sc i\kern-.025em b}\kern-.08em
    T\kern-.1667em\lower.7ex\hbox{E}\kern-.125emX}}
\begin{document}

\title{Towards Perturbation-Induced Static Pivoting \\on GPU-Based Linear Solvers\\
}

\author{\IEEEauthorblockN{Samuel Chevalier}
\IEEEauthorblockA{\textit{Department of Electrical Engineering} \\
\textit{University of Vermont}\\
Burlington, Vermont, USA\\
schevali@uvm.edu}
\and
\IEEEauthorblockN{Robert B Parker}
\IEEEauthorblockA{\textit{Center for Nonlinear Studies} \\
\textit{Los Alamos National Laboratory}\\
Los Alamos, New Mexico, USA \\
rbparker@lanl.gov}
}

\maketitle

\begin{abstract}
Linear system solving is a key tool for computational power system studies{, e.g., optimal power flow, transmission switching, or unit commitment}. CPU-based linear system solver speeds, however, have saturated in recent years. Emerging research shows that GPU-based linear system solvers are beginning to achieve notable speedup over CPU-based alternatives in some applications. Due to the architecture of GPU memory access, numerical pivoting represents the new bottleneck which prevents GPU-based solvers from running even faster. Accordingly, this paper proposes a matrix perturbation-based method to induce \textit{static} pivoting. Using this approach, a series of perturbed, well-conditioned, pivot-free linear systems are solved in parallel on GPUs. Matrix expansion routines are then used to linearly combine the results, and the true solution is recovered to an arbitrarily high degree of theoretical accuracy. We showcase the validity of our approach on distributed-slack AC power flow solve iterations associated with the PGLib 300-bus test case.

\end{abstract}

\begin{IEEEkeywords}
ACOPF, Gaussian elimination, linear system solver, Neumann expansion, GPU
\end{IEEEkeywords}

\section{Introduction}
Linear system solving is at the heart of numerical computation. Within the field of power systems, a variety of estimation, load flow, and optimization routines rely on fast and robust linear system solvers; such solvers must be able to quickly solve very large systems of equations containing, e.g., poorly conditioned Hessian, Jacobian, or covariance matrices. The most dominant CPU-based linear system solvers are well known in the power flow community: the most commonly used include UMFPACK~\cite{davis1995umfpack}, SuperLU~\cite{li2005overview}, KLU~\cite{davis2010algorithm}, 
MA57~\cite{duff2004ma57}, PARDISO~\cite{Schenk2011}, STUMPACK~\cite{ghysels2016efficient}, among several others. 

A group of CPU-based linear system solvers were benchmarked against emerging GPU-based alternatives in~\cite{SWIRYDOWICZ2022102870}. In this work, the authors focused on solving linear KKT systems resulting from interior point iterations; the associated linear systems were very poorly conditioned (i.e., iterative methods were generally unsuitable). For GPU-based solvers, the authors tested CUDA's cuSOLVER~\cite{cusolver} and SPRAL-SSIDS. In 2021, the authors wrote, ``None of the tested packages delivered significant GPU acceleration for our test cases."

Since then, however, GPU-based linear system solvers have accelerated, and specialty tools have arisen to take advantage of their particular capabilities. To accelerate ACOPF solves, authors in \cite{shin2023accelerating} used a condensed-space interior-point method (IPM) with relaxed equality constraints. The resulting approximate KKT matrix could be factored without numerical pivoting, allowing the exploitation of GPU acceleration. The implementations are available through \btt{MadNLP.jl} and \btt{ExaModels.jl}. Also for ACOPF applications, the authors in \cite{swirydowicz2024gpu} recently constructed CPU-GPU hybrid approaches in both cuSOLVER libraries (for use with NVIDIA GPUs) and the open-source Ginkgo libraries~\cite{Anzt_Ginkgo_A_Modern_2022} (for use with AMD GPUs). Through careful permutation, refactorization, and refinement procedures, both specialized solvers were able to outperform generic CPU and GPU solvers. GMRES-based iterative refinement was employed to clean up solutions, and pivoting was identified as a primary challenge.

In this paper, we tackle one of the new emerging bottlenecks in the world of GPU-accelerated linear system solvers: pivoting. As will be reviewed in Sec.~\ref{sec: background}, pivoting is necessary to maintain numerical stability and prevent division by $\sim\!0$. While numerically helpful, pivoting potentially forces the GPU to access large amounts of memory from a variety of distant threads, slowing down Gaussian elimination considerably. As stated in~\cite{swirydowicz2024gpu}, ``...pivoting on the GPU is prohibitively expensive, and avoiding pivoting is paramount for GPU speedups."

In order to avoid pivoting more generally, perturbation-based methods have been proposed~\cite{duff2007towards}. Such methods perturb the diagonal of original matrix $A$ by some small matrix $D$ in order to compute $LU=A+D$. So-called iterative refinement (based on iterative methods such as CG or GMRES) is then used to refine the perturbed solution. The SuperLU\_DIST package~\cite{SuperLU_DIST} is built around this ``perturb then refine" strategy. Iterative refinement, however, can be very slow to converge in the context of a poorly conditioned linear systems~\cite{swirydowicz2024gpu}, and it is {inherently} serial (i.e., iterative).

In this paper, we build upon the perturbation-based approach. However, we offer a new perspective on its application: rather than iteratively refining a single perturbed solution, we use the computational might of GPUs to solve a whole series of differently perturbed systems in parallel through ``batch solving". Then, we leverage the sequence of perturbed solutions to explicitly reconstruct the original, desired solution to an arbitrary degree of theoretical accuracy. This approach is computationally heavy, but most of the computation can be performed in parallel. Importantly, each of the parallel solves can be efficiently performed with static pivoting on GPUs.

This paper focuses on validating the mathematical underpinnings of our perturbation-based approach.
An important sub-task associated with our approach consists of choosing the perturbation matrices which will reliably induce static pivoting. This task is left for future work, and it is replaced by a heuristic which works well in practice. Our paper primarily focuses on the following question: given a perturbation that will induce static pivoting, ($i$) what set of linear systems need to be parallel-solved to reconstruct the desired solution, and ($ii$) what linear combination of perturbed solutions will lead to an optimal reconstruction? {As grid complexity increases without bound, GPUs can be a key technology for both improving performance and expanding capabilities in grid operation and planning. Effective utilization of GPUs, therefore, could help unlock a new frontier of computational possibilities.}

The specific contributions of this paper follow:

\begin{enumerate}
    \item We use a Neumann series matrix expansion to find the linear combination of perturbed system solutions which can optimally reconstruct the desired solution of an original linear system of equations.
    \item We prove that a linear combination of $n$ parallel linear system solutions can theoretically reconstruct the desired (i.e., true) solution with ${\mathcal O}(\epsilon^n)$ accuracy.
    \item We show that the optimal combination can be computed by inverting a transposed Vandermonde matrix.
\end{enumerate}

\section{Background}\label{sec: background}
In this section, we present the matrix expansion (i.e., Neumann expansion) that we build our perturbation-based solver around, and we briefly review pivoting.

\subsection{Taylor series expansion of an inverted scalar function}
Consider some nonzero scalar $a\in {\mathbb R}^1$ and its associated inverse $\frac{1}{a}$. The Taylor series expansion of it \textit{perturbed} inverse, $\frac{1}{a+\epsilon}$, is familiarly given by
\begin{subequations}
\begin{align}\label{eq:scalar_1}
\frac{1}{a+\epsilon} & =a^{-1}-\frac{1}{1!}a^{-2}\epsilon+\frac{1}{2!}a^{-3}\epsilon^{2}-\frac{6}{3!}a^{-4}\epsilon^{3}+\cdots\\
 & =\sum_{i=0}^{\infty}(-1)^{i}a^{-(i+1)}\epsilon^{i},\label{eq:scalar_2}
\end{align}
\end{subequations}
with associated region of convergence $\epsilon<1$: this series expansion is therefore only valid when $\epsilon$ is small.

\subsection{Neumann expansion of a perturbed linear system}
We now consider a square, invertible matrix $A\in{\mathbb R}^{n\times n}$ with associated matrix inverse $A^{-1}$. We perturb $A$ with some diagonal matrix\footnote{Matrix $D$ does not need to be diagonal; the Neumann series expansion is valid for any dense perturbation matrix. We use a diagonal matrix here, since this is historically the most useful candidate for inducing static pivoting~\cite{duff2007towards}.} $D$ which is scaled by a small parameter $\epsilon$. The Neumann series expansion~\cite{el2011variation} of the inverse of this perturbation is given as
\begin{subequations}
\begin{align}(A+\epsilon D)^{-1}= & A^{-1}-\left(A^{-1}\epsilon D\right)^{1}A^{-1}+\left(A^{-1}\epsilon D\right)^{2}A^{-1}\nonumber\\
 & -\left(A^{-1}\epsilon D\right)^{3}A^{-1}+\cdots\label{eq:neumann_1}\\
= & \sum_{i=0}^{\infty}(-1)^{i}\left(A^{-1}\epsilon D\right)^{i}A^{-1}\label{eq:neumann_2}
\end{align}
\end{subequations}
with associated region of convergence $\epsilon<1/\rho(A^{-1}D)$, where $\rho(\cdot)$ is the spectral radius operator. Notably, \eqref{eq:neumann_1}, \eqref{eq:neumann_2} are higher dimension analogies of \eqref{eq:scalar_1}, \eqref{eq:scalar_2}.

\subsection{Pivoting sequence in a linear system solve}
Gaussian elimination of linear systems relies on sequentially dividing rows of a matrix by a leading diagonal entrant. If this value is close to 0, then the row reduction algorithm may have to \textit{pivot}, i.e., exchange matrix rows. The exchanging of matrix rows can be very slow on GPU architecture. When insitu pivoting can be avoided entirely, ``static pivoting" (where a pivot/permutation sequence is either computed once, and perpetually reused, or otherwise known apriori) can be exploited, and GPU-based matrix factorization can run much faster~\cite{swirydowicz2024gpu}. In this paper, we use a perturbation-based approach to induce static pivoting. That is, we perturb a given matrix $A$ by some small matrix. We note that for any given matrix, there always exists some diagonal perturbation matrix which would induce static pivoting. The resulting solution, however, would represent the solution of the \textit{perturbed} system, rather than the original system. In this paper, we solve for a series of perturbed solutions (under static pivoting) in parallel and reconstruct the original solution to a desired degree of accuracy.

\section{Linear Solver with Static Pivoting}
We consider the linear system
\begin{align}
Ax=b
\end{align}
with $A$, $b$ known and $x$ unknown. Without loss of generality, we assume this system is scaled such that $\left\Vert b\right\Vert=1$. We denote the solution of this nonsingular system as $x^{\star}\triangleq A^{-1}b$. Notably, explicit matrix inversion is never actually used to solve large linear systems. A primary computational challenges associated with solving this system on GPUs is pivoting.

\subsection{Static pivoting}
In order to induce a \textit{static} pivoting sequence, we sum a diagonal perturbation matrix $D$ with matrix $A$. The purpose of adding this diagonal perturbation is to ultimately enable static pivoting. Without loss of generality, we assume the entries of $D$ are scaled such that $\left\Vert D\right\Vert _{1}=1$, i.e., the matrix 1 norm is equal to 1. Of course, even though the perturbed problem $(A+D)x=b$ is ``easier" to solve (thanks to static pivoting), its solution may be far from the true solution $x^{\star}$. To overcome this challenge, we solve a group of differently perturbed linear systems, and then we sum the solutions together is a way which approximates $x^{\star}$ with a high degree of accuracy. 

In order to find the set of perturbed solutions which can reconstruct $x^{\star}$, we take a Neumann expansion of the perturbed systems. Generally, the series expansion of $(A+D)x=b$ may not converge, so we modify this system in two ways:

\begin{itemize}
    \item We multiply $D$ by a sufficiently small scalar $\epsilon$ to ensure the series expansion will converge;
    \item We further multiply $\epsilon D$ by a scalar factor $\alpha$ which will allow us to solve for various scaled expansions.
\end{itemize}

Using these perturbed expansion, we define $x_{\alpha}^{+}$ and $x_{\alpha}^{-}$ as the solutions to the following perturbed linear systems:
\begin{align}
x_{\alpha}^{+} & \triangleq(A+\alpha\cdot\epsilon D)^{-1}b\label{eq: xap}\\
x_{\alpha}^{-} & \triangleq(A-\alpha\cdot\epsilon D)^{-1}b.\label{eq: xam}
\end{align}
\begin{remark}\label{rem_avg}
    Taking the average of \eqref{eq: xap} and \eqref{eq: xam} yields a series solution with only even power of $\epsilon$.
\end{remark}

\begin{figure*}[b]
\begin{align}\label{eq: sum_plus_and_minus} 
\begin{array}{crcccccccc}
 & \tfrac{1}{2}(A+\alpha\cdot\epsilon D)^{-1}= & \tfrac{1}{2}A^{-1} & - & \frac{\alpha}{2}\left(A^{-1}\epsilon D\right)^{1}A^{-1} & + & \frac{\alpha^{2}}{2}\left(A^{-1}\epsilon D\right)^{2}A^{-1} & - & \frac{\alpha^{3}}{2}\left(A^{-1}\epsilon D\right)^{3}A^{-1} & +\;\cdots\\[0.1cm]
+ & \tfrac{1}{2}(A-\alpha\cdot\epsilon D)^{-1}= & \tfrac{1}{2}A^{-1} & + & \frac{\alpha}{2}\left(A^{-1}\epsilon D\right)^{1}A^{-1} & + & \frac{\alpha^{2}}{2}\left(A^{-1}\epsilon D\right)^{2}A^{-1} & + & \frac{\alpha^{3}}{2}\left(A^{-1}\epsilon D\right)^{3}A^{-1} & +\;\cdots\\[0.1cm]
\hline  & = & A^{-1} &  &  & + & \alpha^{2}\left(A^{-1}\epsilon D\right)^{2}A^{-1} &  &  & +\;\cdots
\end{array}
\end{align}
\end{figure*}

The validity of Remark \ref{rem_avg} is shown explicitly in \eqref{eq: sum_plus_and_minus}, where all odd powers of $\epsilon$ cancel out. Taking the average of $x_{\alpha}^{+}$ and $x_{\alpha}^{-}$ can be thought of as an approximation \textit{correction}, since this average offers a closer approximation to $x^{\star}$ than either $x_{\alpha}^{+}$ or $x_{\alpha}^{-}$ do individually. In this paper, we propose solving for further refined approximations by including other perturbed solutions. To build up these perturbations, we define the average of \eqref{eq: xap}-\eqref{eq: xam} for $\alpha=1,2,3,\dots,m$ as $\hat{x}_{1}$, $\hat{x}_{2}$, $\hat{x}_{3}$, $\dots,$ $\hat{x}_{m}$  in \eqref{eq: avg_hats}. We refer to these as ``averaged perturbation solutions", and their weighted sum will yield a close approximation for $x^{\star}$. Notably, the use of integer values for $\alpha$ is a modeling choice; smaller decimal values ($\alpha=1.0,1.1,1.2,\dots$) could alternatively be used, but this could cause numerical imprecision.

Using the averaged perturbation solutions from \eqref{eq: avg_hats}, we present the following main result. This result assumes a short sequence of perturbations $\alpha=1,\dots,m$, where $\epsilon_c\triangleq m \cdot \epsilon$ is within the region of convergence for \eqref{eq:neumann_2}.

\begin{theorem}\label{eq: thm_error}
There exists a linear combination of averaged perturbation solutions $\hat{x}_{1}$, $\hat{x}_{2}$, $\hat{x}_{3}$, $\dots$, $\hat{x}_{m}$ which can recover $x^{\star}$ with ${\mathcal O}(\epsilon^{2m})$ accuracy.
\begin{proof}
With slight abuse of notation, rewrite \eqref{eq: avg_hats} via the following matrix-vector product $\Gamma v$:
\begin{align}\label{eq: gam_v}
\left[\!\!\begin{array}{c}
\hat{x}_{1}\\
\hat{x}_{2}\\
\hat{x}_{3}\\
\vdots\\
\hat{x}_{m}
\end{array}\!\!\right]\!=\!\underbrace{\left[\!\begin{array}{cccc}
1^{0} & 1^{2} & 1^{4} & \cdots\\
2^{0} & 2^{2} & 2^{4} & \cdots\\
3^{0} & 3^{2} & 3^{4} & \cdots\\
 & \vdots\\
m^{0} & m^{2} & m^{4} & \cdots
\end{array}\!\right]}_{\Gamma}\!\underbrace{\left[\!\!\begin{array}{c}
\left(A^{-1}\epsilon D\right)^{0}A^{-1}b\\
\left(A^{-1}\epsilon D\right)^{2}A^{-1}b\\
\left(A^{-1}\epsilon D\right)^{4}A^{-1}b\\
\vdots
\end{array}\!\!\right]}_{v}\!\!.
\end{align}
Since $\Gamma$ is a Vandermonde matrix, its rows are necessarily linearly independent~\cite{Horn:1990}. Therefore, a linear combination of $m$ rows can eliminate $m-1$ RHS terms (i.e., terms attached to various powers of $\epsilon$) via row-reduction techniques. Assume the goal is to eliminate the lowest non-zero power of $\epsilon$. Since the RHS terms increment by powers of $\epsilon^2$, and since the first RHS term $A^{-1}b$ is always retained, the highest remaining power of $\epsilon$ after the elimination is of power $2(m-1)+2=2m$.
\end{proof}
\end{theorem}

We note that a series of $2m=n$ linear systems must be solved to produce the $m$ averaged expressions in \eqref{eq: avg_hats}. Therefore, Theorem \ref{eq: thm_error} implies the error scales via ${\mathcal O}(\epsilon^{2m})={\mathcal O}(\epsilon^{n})$, where $n$ is the total number of linear system solves. We now offer two examples of this theorem for $m=2$ and $m=3$.

\textit{Example 1.} Consider $m=2$, where the best approximation to $x^{\star}$ is computed as ${\hat x}^{\star}$ via the following linear combination:
\begin{subequations}
\begin{align}
\hat{x}^{\star} =&\frac{\hat{x}_{2}-4\hat{x}_{1}}{-3}\label{eq: lin_comb1}\\
  =&\left(A^{-1}-4\left(A^{-1}\epsilon D\right)^{4}A^{-1}\right.\nonumber\\
 &\qquad\qquad \left.-20\left(A^{-1}\epsilon D\right)^{6}A^{-1}+\cdots\right)b.
\end{align}
\end{subequations}
The associated error $e=\left\Vert x^{\star}-\hat{x}^{\star}\right\Vert$ is given by
\begin{align}\label{eq: m2_error}
e & =\left\Vert x^{\star}-\hat{x}^{\star}\right\Vert \\
 & =\left\Vert A^{-1}b-\left(A^{-1}+4\left(A^{-1}\epsilon D\right)^{4}A^{-1}\right. \right.\nonumber \\
 &\qquad\qquad\qquad\qquad\left.\left.+20\left(A^{-1}\epsilon D\right)^{6}A^{-1}+\cdots\right)b\right\Vert\nonumber  \\
 & =\left\Vert \left(4\left(A^{-1}\epsilon D\right)^{4}A^{-1}+20\left(A^{-1}\epsilon\nonumber  D\right)^{6}A^{-1}+\cdots\right)b\right\Vert \\
 & \le\left\Vert \left(4\left(A^{-1}\epsilon D\right)^{4}A^{-1}+20\left(A^{-1}\epsilon\nonumber D\right)^{6}A^{-1}+\cdots\right)\right\Vert \left\Vert b\right\Vert \\
 & =\mathcal{O}(\epsilon^{4}),\nonumber
\end{align}
since $\left\Vert b\right\Vert=1$, by previous assumption.\hfill\qedsymbol{}

\textit{Example 2: $m=3$.} Consider the $m=3$ case, where the best approximation to $x^{\star}$ is computed as ${\hat x}^{\star}$ via the following linear combinations. First, we define two intermediate corrections, each of which have no $\epsilon^2$ term:
\begin{align}\label{eq: x21}
\hat{x}_{21}^{*} & =\frac{\hat{x}_{2}-4\hat{x}_{1}}{-3}\\
\hat{x}_{31}^{*} & =\frac{\hat{x}_{3}-9\hat{x}_{1}}{-8}.\label{eq: x31}
\end{align}
Next, we build ${\hat x}^{\star}$ as a linear combination of \eqref{eq: x21}-\eqref{eq: x31}:
\begin{subequations}
\begin{align}
\hat{x}^{\star} & =\hat{x}_{31}^{*}-\frac{9}{4}\hat{x}_{21}^{*}\\
 & =A^{-1}-36\left(A^{-1}\epsilon D\right)^{6}A^{-1}-\cdots,
\end{align}
\end{subequations}
which has approximation error ${\mathcal O}(\epsilon^6)$ (steps omitted). \hfill\qedsymbol{}

The optimal linear combination identified in \eqref{eq: lin_comb1}, can be readily identified by inspection of \eqref{eq: xh1}-\eqref{eq: xh2}: this unique combination of vectors will ``cancel out" $\epsilon^2$ terms, and it will retain an unscaled $A^{-1}$ term. However, through the following remark, we can generalize the identification of these linear combinations. In this remark, we define $\Gamma_m$ as the left-most $m \times m$ submatrix of $\Gamma$ in \eqref{eq: gam_v}. For example, $\Gamma_3$ is given as
\begin{align}
\Gamma_{3}=\left[\begin{array}{ccc}
1^{0} & 1^{2} & 1^{4}\\
2^{0} & 2^{2} & 2^{4}\\
3^{0} & 3^{2} & 3^{4}
\end{array}\right].
\end{align}

\begin{remark}\label{rem_vinv}
    Given a set of $m$ vectors $\hat{x}_{1}$, $\ldots$, $\hat{x}_{m}$, the optimal approximation $\hat{x}^{\star}$ to true solution ${x}^{\star}$ is given as
\begin{align}\label{eq: x_star_vand}
\hat{x}^{\star}=\left[\!\begin{array}{cccc}
\hat{x}_{1} & \hat{x}_{2} & \cdots & \hat{x}_{m}\end{array}\!\right]\left(\Gamma_{m}^{T}\right)^{-1}\left(\left[1\;0\;0\,\cdots\,0\right]^T\right)\!.
\end{align}
\end{remark}

\begin{figure*}[b]
\begin{subequations}\label{eq: avg_hats}
\begin{align}
\hat{x}_{1}\triangleq\tfrac{1}{2}\left(x_{\alpha=1}^{+}+x_{\alpha=1}^{-}\right)=\; & \left(A^{-1}\;+\;1\left(A^{-1}\epsilon D\right)^{2}A^{-1}\;+\;01\left(A^{-1}\epsilon D\right)^{4}A^{-1}\;+\;001\left(A^{-1}\epsilon D\right)^{6}A^{-1}\;+\;\cdots\right)b\label{eq: xh1}\\[-3pt]
\hat{x}_{2}\triangleq\tfrac{1}{2}\left(x_{\alpha=2}^{+}+x_{\alpha=2}^{-}\right)=\; & \left(A^{-1}\;+\;4\left(A^{-1}\epsilon D\right)^{2}A^{-1}\;+\;16\left(A^{-1}\epsilon D\right)^{4}A^{-1}\;+\;064\left(A^{-1}\epsilon D\right)^{6}A^{-1}\;+\;\cdots\right)b\label{eq: xh2}\\[-3pt]
\hat{x}_{3}\triangleq\tfrac{1}{2}\left(x_{\alpha=3}^{+}+x_{\alpha=3}^{-}\right)=\; & \left(A^{-1}\;+\;9\left(A^{-1}\epsilon D\right)^{2}A^{-1}\;+\;81\left(A^{-1}\epsilon D\right)^{4}A^{-1}\;+\;729\left(A^{-1}\epsilon D\right)^{6}A^{-1}\;+\;\cdots\right)b\\[-7pt]
\vdots\;\; & \nonumber \\[-7pt]
\hat{x}_{m}\triangleq\,\tfrac{1}{2}\left(x_{\alpha=m}^{+}+x_{\alpha=m}^{-}\right)=\; & \left(A^{-1}\;+\;m^{2}\left(A^{-1}\epsilon D\right)^{2}A^{-1}\;+\;m^{4}\left(A^{-1}\epsilon D\right)^{4}A^{-1}\;+\;\;\;m^{6}\left(A^{-1}\epsilon D\right)^{6}A^{-1}\,+\,\cdots\right)b\label{eq: xhm}
\end{align}
\end{subequations}
\end{figure*}
We prove remark \ref{rem_vinv} via construction. Assume the optimal approximation is written as the following linear sum for a set of constant $\beta_i$ values:
\begin{align}\label{eq: x_star}
    \hat{x}^{\star} =\sum_i\beta_{i}\hat{x}_{i}.
\end{align}
Now, scale each \eqref{eq: xh1}-\eqref{eq: xhm} by its associated $\beta_{i}$ constant, sum the expressions, and collect $\left(A^{-1}\epsilon D\right)^{k}A^{-1}b$ terms. We denote $\gamma_k$ as the coefficient in front of the collection of these terms. For example, for the $k=0$ case, we have
\begin{align}
\beta_{1}(A^{-1}b)+\beta_{2}(A^{-1}b)+\cdots=\underbrace{\left(\beta_{1}+\beta_{2}+\cdots\right)}_{\gamma_{0}}A^{-1}b.
\end{align}
Dropping the terms $\left(A^{-1}\epsilon D\right)^{k}A^{-1}b$ (which show up like invertible diagonal matrices on either side of the following expression), the $\gamma$ terms may be computed via,
\begin{align}\label{eq: gam}
\underbrace{\left[\begin{array}{cccc}
1^{0} & 2^{0} & 3^{0}\\
1^{2} & 2^{2} & 3^{2} & \cdots\\
1^{4} & 2^{4} & 3^{4}\\
 & \vdots &  & \ddots
\end{array}\right]}_{\Gamma_{m}^{T}}\left[\begin{array}{c}
\beta_{1}\\
\beta_{2}\\
\beta_{3}\\
\vdots
\end{array}\right]=\left[\begin{array}{c}
\gamma_{0}\\
\gamma_{2}\\
\gamma_{4}\\
\vdots
\end{array}\right]
\end{align}
where $\Gamma_{m}^{T}$ is the transposed Vandermonde matrix from \eqref{eq: gam_v}. We also assume $\left(A^{-1}\epsilon D\right)^{k}A^{-1}b$ terms with $k>2(m-1)$ have been dropped, so that $\Gamma_{m}^{T}$ is a square matrix (these dropped higher order terms can only be canceled with a larger choice of $m$). Since the right hand side vector in \eqref{eq: gam} represents the relative sizes of grouped powers of $\epsilon$, we wish to determine the set of $\beta$ values which will give us
\begin{align}
\left[\gamma_{0}\;\gamma_{2}\;\gamma_{4}\;\cdots\right]^T=\left[1\;0\;0\;\cdots\right]^T,
\end{align}
i.e., the leading power with $A^{-1}$ has a unity scale, and all other (non-neglected) higher order terms are canceled out (corresponding to a $\gamma_i=0$). Once these $\beta_i$ values are solved for (by inverting $\Gamma_{m}^{T}$), the optimal solution in \eqref{eq: x_star_vand} is constructed by summing the appropriately scaled ${\hat x}_i$ vectors.

\textit{Example 3.} Consider the $m=2$ case, where
\begin{align}
\Gamma_{2}^{T}=\left[\begin{array}{cc}
1^{0} & 1^{2}\\
2^{0} & 2^{2}
\end{array}\right].
\end{align}
The associated system solution is 
\begin{align}
\left[\begin{array}{c}
\beta_{1}\\
\beta_{2}
\end{array}\right]=\left(\Gamma_{2}^{T}\right)^{-1}\left[\begin{array}{c}
1\\
0
\end{array}\right]=\frac{1}{3}\left[\begin{array}{c}
4\\
-1
\end{array}\right],
\end{align}
where $\beta_1$ and $\beta_2$ are the same set of $\beta$ values implicitly used in \eqref{eq: lin_comb1}, yielding an eventual error of $\mathcal{O}(\epsilon^4)$ in \eqref{eq: m2_error}.

\subsection{Methodology Summary}
The proposed mythology is summarized in Alg. \ref{algo:pistol}. We assume the perturbation matrix $D$ is known apriori. 


\begin{algorithm}
\caption{Perturbation-Induced Static Pivoting.}\label{algo:pistol}

{\small

\begin{algorithmic}[1]

\State Parallel perturb a given matrix $A$ by $\pm \alpha \cdot \epsilon D$ for a chosen $\alpha$ sequence, depending on the desired level of solution precision

\State Batch factorize and solve each of the perturbed systems

\State Construct vectors ${\hat x}_i$ from \eqref{eq: avg_hats}

\State Linear combine ${\hat x}_i$ via \eqref{eq: x_star_vand} to reconstruct optimal approximation

\State Optionally, test the solution error

\State \Return Approximated solution ${\hat x}^\star\approx {x}^\star$
\end{algorithmic}}
\end{algorithm}

\section{Test Results}
In this section, we demonstrate the validity of the proposed formulations, {and} we provide motivating timing analysis. We use CUDA's cuSOLVER library, accessed through \btt{CUDA.jl}, to perform the batched GPU-based linear system solves.

\begin{figure}
\centering
\includegraphics[width=0.95\columnwidth]{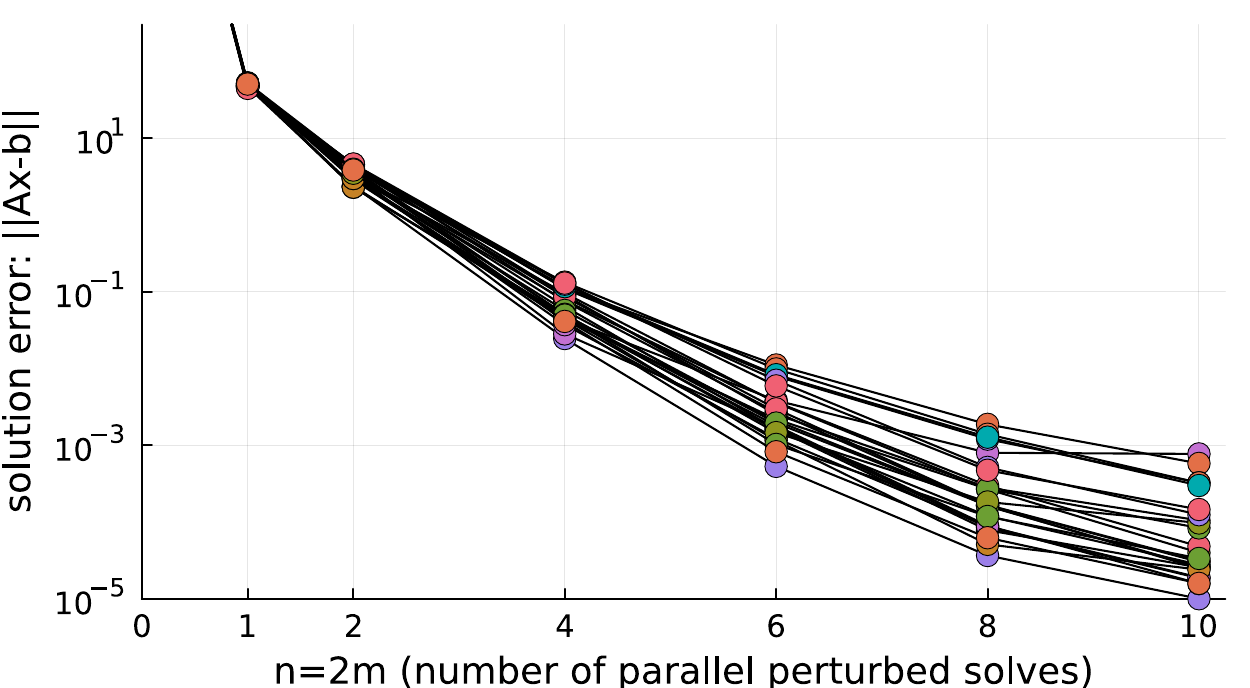}
\caption{Log scale plot of the solution approximation error $\left\Vert J\hat{x}^{\star}-b\right\Vert $ for increasing values of $n$. As the number of parallel perturbed solves increase, the accuracy of solution \eqref{eq: x_star} increases exponentially, until numerical saturation is reached (around $10^{-5}$). This numerical saturation is driven partially by the ill conditioning of the Vandermonde matrix itself. 25 trials were considered with Jacobians evaluated under different voltage starting values.}\label{fig:errors}
\end{figure}

In this test, we solved a distributed slack~\cite{Dhople:2020} AC power flow problem on the PGLib 300-bus test case using Newton-Raphson. A linear system solver is needed at each step of Newton-Raphson to solve $J\Delta x = b$, where $J\in{\mathbb R}^{531\times 531}$ is an appropriately reduced distributed slack power flow Jacobian; we evaluate $J$ with many different $V$, $\theta$ initializations (i.e., to simulate different directional steps). In our experiment, we compared GPU-based linear system solves using ($i$) a no-pivoting direct solve, versus ($ii$) a no-pivoting solve with our proposed approach. Results are summarized below:
\begin{itemize}
\item Due to the structure of the distributed slack power flow Jacobian, the GPU-based linear system solve with no pivoting returned a vector filled with \btt{NaN} and \btt{Inf} values. A leading 0 was always encountered on a diagonal, so no solution could be found. As benchmark, we additionally applied the Hopcroft-Karp algorithm (which permutes a matrix by moving non-zeros onto the diagonal). This algorithm ran in 1.4 ms, and the resulting (pivot-free) linear system solve still resulted in \btt{NaN}/\btt{Inf} values.

\item We then applied Alg.~\ref{algo:pistol} for various numbers of parallel linear system solves ($n$). Notably, we were testing on a case where a pivot-free solve returned a meaningless solution. Our method, however, enabled static pivoting. Fig.~\ref{fig:errors} shows that an increasing number of parallel perturbed solutions can drive approximation error to $\sim10^{-5}$.
\end{itemize}

\begin{figure}
\centering
\includegraphics[width=1.0\columnwidth]{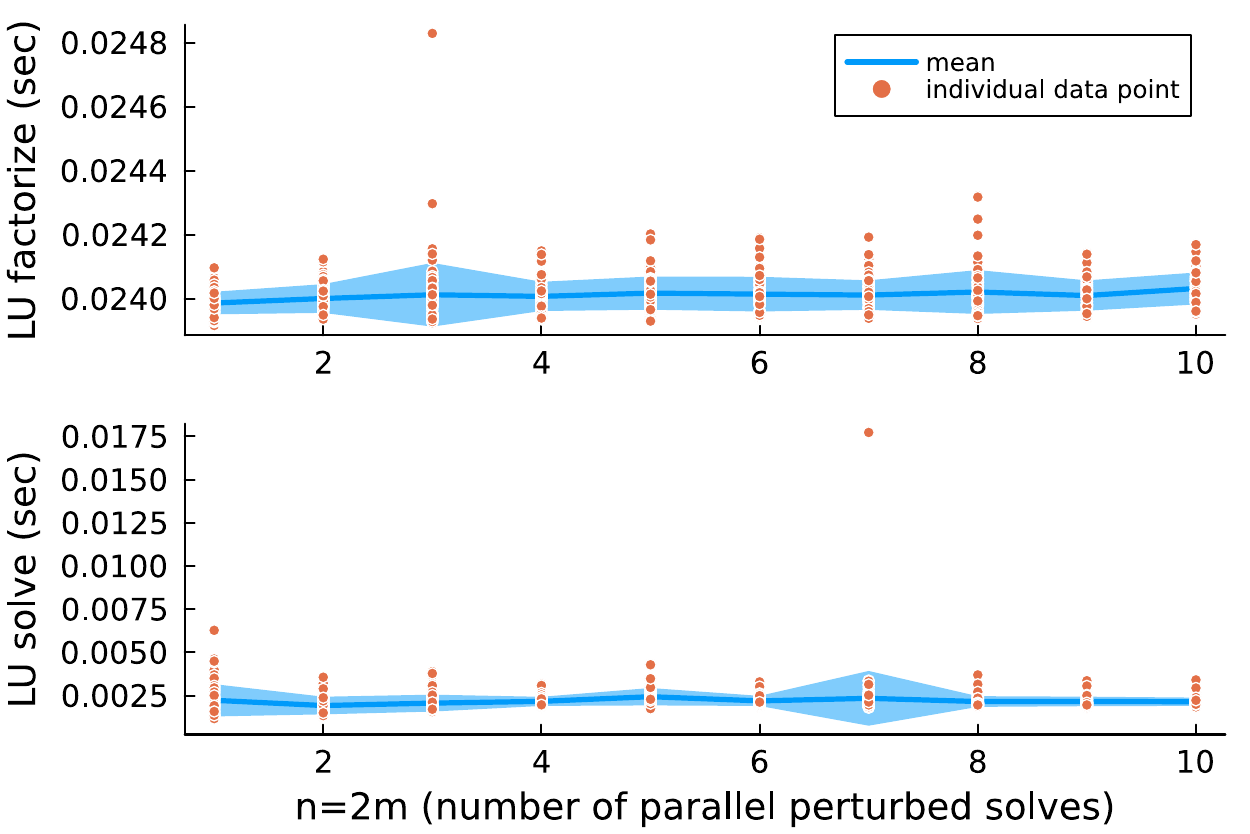}
\caption{Time spent performing LU factorization (top) vs forward-backward solving (bottom) over 100 trials (for each $n$). The GPU can perform $n=1$ to $n=10$ parallel LU factorization in approximately the same amount of time.}\label{fig:timing}
\end{figure}

We also performed timing analysis of the parallel linear system solves, timed with \btt{CUDA.@elapsed}, as demonstrated in Fig.~\ref{fig:timing}. The top panel times batched LU factorization via the function \btt{CUBLAS.getrf\_batched!($\cdot$)}, and the bottom panel times batched forward-elimination/back-substitution via successive calls to \btt{CUBLAS.trsm\_batched!($\cdot$)}. Notably, GPU solve time is \textit{invariant} to the number of parallel batched factorizations and solves. Said differently, the best solutions and the worst solutions illustrated Fig~\ref{fig:errors} require approximately the same amount of wall-clock time to compute. Following are a few notes on practical implementation:
\begin{itemize}
\item Even though $J$ is sparse, there is no sparse \textit{batched} LU solver available in \btt{CUDA.jl} yet, so \btt{getrf\_batched!} and \btt{trsm\_batched!} are dense operators. Thus, timing analysis of Fig.~\ref{fig:timing} is provided to show relative timing requirements for different levels of batching.

\item In our test, we used $\epsilon=2\times10^{-3}$, and all entries of the perturbation matrix were chosen to be a normally distributed. Interestingly, normally distributed perturbations tended to yield lower error than a scaled identity matrix.

\item Compared to factorization and solve time, summing 10 perturbed solutions (for $n=10$ case) was trivial ($\sim 5 \, \mu s$).

\end{itemize}

\section{Conclusions}
This paper has offered a new perturbation-based framework for inducing static pivoting. Once fully developed, our method can help to accelerate GPU-based linear system solvers by restricting the amount of data they must call from distant threads. We demonstrated the mathematical validity of our proposed scheme on a 300-bus distributed slack power flow solve, driving an infinite degree of step error (in the pivot-free, perturbation-free solution) to $e=10^{-5}$ (in the case of 10 perturbed linear systems solved in parallel). {GPU-based matrix factorization could enable thousands of nonlinear power systems models to be solved in parallel on a single GPU, greatly enhancing the branch-and-bound convergence time for applications like the AC Unit Commitment, or enable the infusion of AC power flow solves within notoriously large production cost modeling problems.}

Further work is needed on several fronts.
First, the perturbation strategy must be rigorously investigated.
The authors observed that the perturbations had a large effect on the error convergence rate.
Despite the success of normally distributed perturbations, more systematic strategies should be investigated.
Second, the methods should be optimally implemented with batched factorization and solve functions which operate on sparse matrices. Finally, the method needs to be tested on larger and more poorly conditioned systems.
\\

{\small\noindent R.P. acknowledges funding from the Center for Nonlinear Studies at Los Alamos National Laboratory. LA-UR-23-33525.}

\bibliographystyle{ieeetr}
\bibliography{references}
\end{document}